\documentclass{egpubl}
\usepackage{STAG2022}
 
\ConferencePaper        
 \WsPaper           

\usepackage[T1]{fontenc}
\usepackage{dfadobe}

\usepackage{amsmath}
\usepackage{amssymb}

\biberVersion
\BibtexOrBiblatex
\usepackage[backend=biber,bibstyle=EG,citestyle=alphabetic,backref=true]{biblatex} 
\electronicVersion
\PrintedOrElectronic

\ifpdf \usepackage[pdftex]{graphicx} \pdfcompresslevel=9
\else \usepackage[dvips]{graphicx} \fi

\usepackage{egweblnk} 

\title{Optimizing Placements of 360$^{\circ}$ Panoramic Cameras in Indoor Environments by Integer Programming}

\author[Syuan-Rong Syu \& Chi-Han Peng]
{\parbox{\textwidth}{\centering Syuan-Rong Syu$^1$ and Chi-Han Peng$^1$\orcid{0000-0002-6823-8029}}
        \\
{\parbox{\textwidth}{\centering National Yang Ming Chiao Tung University$^1$}
}
}

\begin{document}

 \teaser{
  \includegraphics[width=\linewidth]{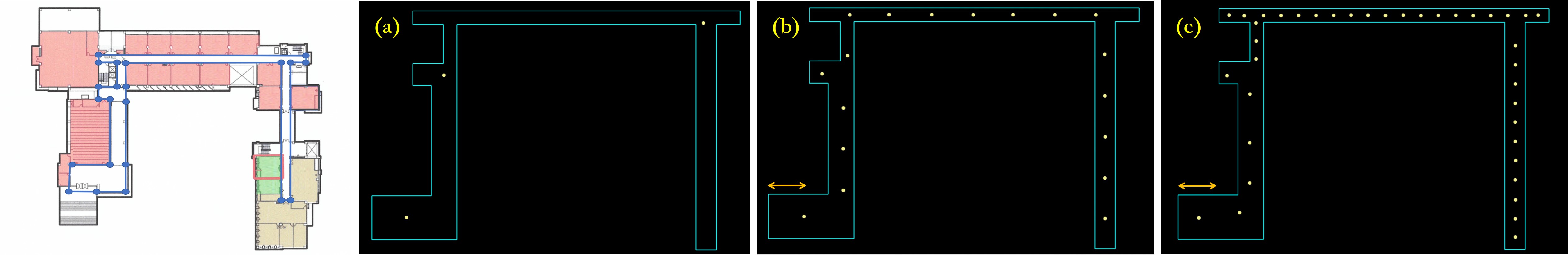}
  \centering
   \caption{Our method finds globally optimal (in terms of numbers of camera placements) solutions to place 360$^{\circ}$ cameras that together completely cover all visible regions of interests (ROIs) of an indoor environment. We model the indoor environment according to the public area of a real-world floor plan of an office building (left), and set the ROIs to be the walls of the area. The three shown solutions ((a) to (c)), are one that has no constraints (i.e., the 360$^{\circ}$ camera is omnidirectional and can see infinitely far away), one that has a maximal range constraint (orange), and one that has both a maximal range and an angle-to-wall constraint (w.r.t. surface normals) of 45 degrees.}
 \label{fig:teaser}
}

\maketitle
\begin{abstract}
We propose a computational approach to find a minimal set of 360$^{\circ}$ camera placements that together sufficiently cover an indoor environment for the building documentation problem in the architecture, engineering, and construction (AEC) industries. Our approach, based on a simple integer programming (IP) problem formulation, solves very efficiently and globally optimally. We conducted a study of using panoramas to capture the appearances of a real-world indoor environment, in which we found that our computed solutions are better than human solutions decided by both non-professional and professional users.


\end{abstract}  

\section{Introduction}

Taking photos to record and document the interiors of a building is an important task in architecture, engineering, and construction (AEC). For examples, during constructions of a building, firms have to document the progress by taking numerous photos, over time, of the interiors of the building or the jobsite~\cite{OpenSpace}. In insurance and restoration, surveyors have to thoroughly document the state of a damaged building for repair cost estimations~\cite{Matterport2022}. Real-estates agents shoot photos of the interiors of houses for sale. Even in retail and hospitality, owners have to showcase the appearances of the indoor spaces to customers. A growing trend is to replace traditional perspective photos by panoramic photos, or "panoramas" in short, shoot by 360$^{\circ}$ cameras. The main reason is that a panoramic photo covers nearby every angles seen from the camera position (except for the top and bottom nadirs), vastly reducing the numbers of photos need to be taken.


Even though 360$^{\circ}$ cameras shoot photos omnidirectionally, to capture an indoor environment, multiple panoramas are still likely needed due to occlusions and ranges of the cameras. Traditionally, photographers decide placements of the 360$^{\circ}$ camera strategically by intuition and simple guidelines (e.g., placing it at the centers of rooms, plus a few more to capture occluded parts). However, such solutions may not be most economical in terms of the numbers of panoramas taken. Worse, some parts of the indoor environment may not be captured at all due to human miscalculations.

In this paper, we propose a optimization-based approach to the 360$^{\circ}$ camera placement problem. In short, we aim to minimize the number of panoramas taken that together would sufficiently capture all visible parts of an indoor environment. We formulate the problem as a linear integer programming (IP) problem, which can be solved globally-optimally and very efficiently by modern off-the-shelf solvers such as Gurobi~\cite{gurobi}. Note that in our IP formulation, the visibility ranges of the 360$^{\circ}$ cameras can be modeled in straightforward manners.

We narrow our scope by making the following assumptions. First, we assume that the indoor environment is largely clutter-free, which means that it is adequate to encode the indoor environment as a 2.5D model (i.e., a 2D floorplan "extruded" vertically by a given height) consisting of planar faces presenting the walls, floors, and ceilings. This assumption makes sense for a variety of cases such as unfurnished houses and the public areas of large buildings. Second, as we focus on solving the camera placement problem, we assume the 2.5D model of the indoor environment is given. For example, pre-modeled according to the floorplan of the building. In other words, our method can be used as a planning tool before the photographer actually go to the jobsite.

To verify the usefulness of our method, we set up a study to simulate the usage of 360$^{\circ}$ cameras for indoor environment documentation. First, we build a 2.5D model of our department according to its floorplan, and then compare our computed solutions versus solutions decided by amateur users and one professional architect. We find that our solution uses fewer panoramas to completely cover the whole indoor environment. In fact, the human solutions are often incorrect as some parts of the environment were not captured. Second, we estimate the visibility ranges of popular 360$^{\circ}$ cameras used in the AEC industry through experiments, and used the estimated ranges in our IP formulation.

Our contributions are summarized as follows:

\begin{itemize}
\item We propose a computational approach to the 360$^{\circ}$ camera placement problem that was previously mainly solved by human intuition. Our IP-based method is simple, solves very efficiently by a modern off-the-shelf solver on a conventional computer, and provides globally-optimal solutions. We also demonstrate that it is straightforward to impose additional constraints, such as the visibility ranges of the cameras, to the IP formulation.
\item We conducted a study of using panoramas to capture the appearances of a real-world indoor environment, in which we compared our computed solution to solutions decided by humans. We show interesting observations of human behaviors, which may be useful for designing computational approaches in the future.
\end{itemize}


\section{Related Work}
\label{sec:related}

We discuss related work in three main topics: 1) common practices and guidelines for 360$^{\circ}$ camera placement in the construction and real-estate industries, 2) automatic camera placement in sensor network design, and 3) visibility in computer graphics.

\subsection{360$^{\circ}$ camera placement guidelines}

In the Zillow Indoor Dataset paper~\cite{ZInD}, a simple guideline for placing 360$^{\circ}$ cameras to shoot virtual tours is described, which only regulated that more panoramas should be taken in bigger rooms. Similar guidelines (i.e., shoot a panorama at every 2 meters apart) are suggested in a popular website post~\cite{Daniel2021}. In the Matterport's guide for construction documentation~\cite{Matterport2017}, they suggest the following steps for taking panoramas in construction sites: 1) first, create a 2D drawing or floorplan of the site to plan the paths in advance, 2) physically mark previous placements of 360$^{\circ}$ cameras for easier revisiting, and 3) shoot panoramas in the middle of the rooms, plus detailed 3D scans around the perimeters.

\subsection{Automatic camera placement}

The problem of automatic camera placement is an important topic in sensor networks design. The main application is the setup of surveillance cameras in large buildings such as malls, airports, and factories~\cite{ghanem2015designing, ERDEM2006156, gao2009constructing} and even urban areas~\cite{7968252}. Usually the task is formulated as an optimization problem in which the user wants to use as few cameras as possible to completely cover the regions-of-interest (ROI) of an environment~\cite{erdem2004optimal, ghanem2015designing}, or alternatively, maximizing the coverage of the ROI with a limited (budgeted) number of cameras~\cite{horster2006optimal,4587515}. Variations of the problem include different ways to model the visibility of the cameras (e.g., conventional directional cameras or omnidirectional cameras~\cite{5152761}), how the environment is encoded (e.g., discretized into a set of grid points or continuously encoded as a set of
2D polygons as in the the classical art gallery problem~\cite{AGP}, and camera-to-camera relationships~\cite{ghanem2015designing, motamedi2017signage}. We refer readers to survey papers discussing coverage problems in sensor network design~\cite{10.1145/1978802.1978811,10.1007/s11263-012-0587-7}.

Mentioned previously, the {\em art gallery problem} (AGP) is a key topic in computational geometry that is similar to our problem. In short, AGP asks for the minimal placements of "guards" (or cameras) that see omnidirectionally and indefinitely away to completely cover a 2D domain modeled by a set of polygons~\cite{AGP}. AGP are either solved discretely (by sampling the 2D domain and possible guard placement locations into a discrete set of points), which can be reduced to instances of the set covering problem~\cite{GHOSH2010718}, or continuously in which sub-optimal~\cite{doi:10.1142/S0218195910003451,BOTTINO20111048} or optimal solutions~\cite{https://doi.org/10.1111/j.1475-3995.2011.00804.x,10.1145/2890491} are found. Our problem in the unconstrained form (for example, problem (a) in Figure~\ref{fig:teaser}) is equivalent to the special case of AGP in which only the perimeters of the 2D domain need to be covered~\cite{BOTTINO20111048}. However, when range or angle-to-wall constraints are applied, the problem is no longer an AGP and existing algorithms are not directly applicable.

We now focus discussions on the discrete case. An important distinction of related methods is how they actually solved the set covering problem, which is a linear binary problem (LBP) and is NP-hard in nature. Earlier methods solve the LBP by divide-and-conquer strategies~\cite{ERDEM2006156,DUNN20061209}, heuristic greedy methods~\cite{10.1145/1178782.1178801}, being relaxed to linear programming (LP) and then solved by the conventional simplex method~\cite{10.1145/1556134.1556140}, or genetic programming~\cite{5403076}. These methods reportedly often took {\em hours} to solve for medium to large problems~\cite{ghanem2015designing}. Later, in 2015, Delbos et al. proposed to relax the binary problem into a convex quadratic one by replacing the binary constraints with box constraints and then solve the problem by an academic solver~\cite{DG05}. They reported to reduce the computation time to seconds for problems with visibility matrices with several millions of entries. In comparison, we formulate the camera placement problem as a simple linear binary programming problem and solve it using a modern off-the-shelf solver (Gurobi). We solve our problems of roughly the same size as in~\cite{DG05} in less than one second.

\subsection{Visibility in computer graphics}

A closely related topic to camera placements is the study of {\em visibility} in computer graphics, i.e., what can be seen in a 2D or 3D scene along a line, from a point, from a line segment, from a polygon, etc. (quoted from the survey by Bittner and Wonka~\cite{Bittner03visibilityin}). Another exhaustive categorization of visibility in 3D is done by Durand et al.~\cite{durand19963d}. A survey about visibility issues for first-person view applications in large scenes is provided in~\cite{1207447}. In our paper, we consider the visibility of an omnidirectional camera in 2D w.r.t. a closed loop of 2D lines that together encode an indoor environment. This is a well studied problem in computer graphics, especially in 2D game developments~\cite{asano1985efficient,Bungiu14}. However, we took additional considerations, including minimal and maximal ranges (i.e., points too close and too far away from the camera are considered non-visible) and angles to the surface normals (i.e., surface points with directional vectors to the camera position that deviate too much from the surface normals are considered non-visible), that are less discussed in related work.

\section{Method}
\label{sec:method}

We begin with problem definitions. Our goal is to find the positions of a set of 360$^{\circ}$ cameras such that all regions of interest (ROIs) of an indoor environment are visible from at least one of the cameras. Recall that we assume the indoor environment is encoded as a 2.5D model in 3D, consisting of a planar floor, a planar ceiling, and several planar walls. Together they form a water-tight 3D model. We assume the direction of gravity is the -z axis. Therefore, the floor and the ceiling are planes with normals aligned to the z axis (i.e., x-y planes at different heights), and the walls are planes with normals orthogonal to the z axis. 

\begin{figure}[t]
  \centering
  \includegraphics[width=0.9\linewidth]{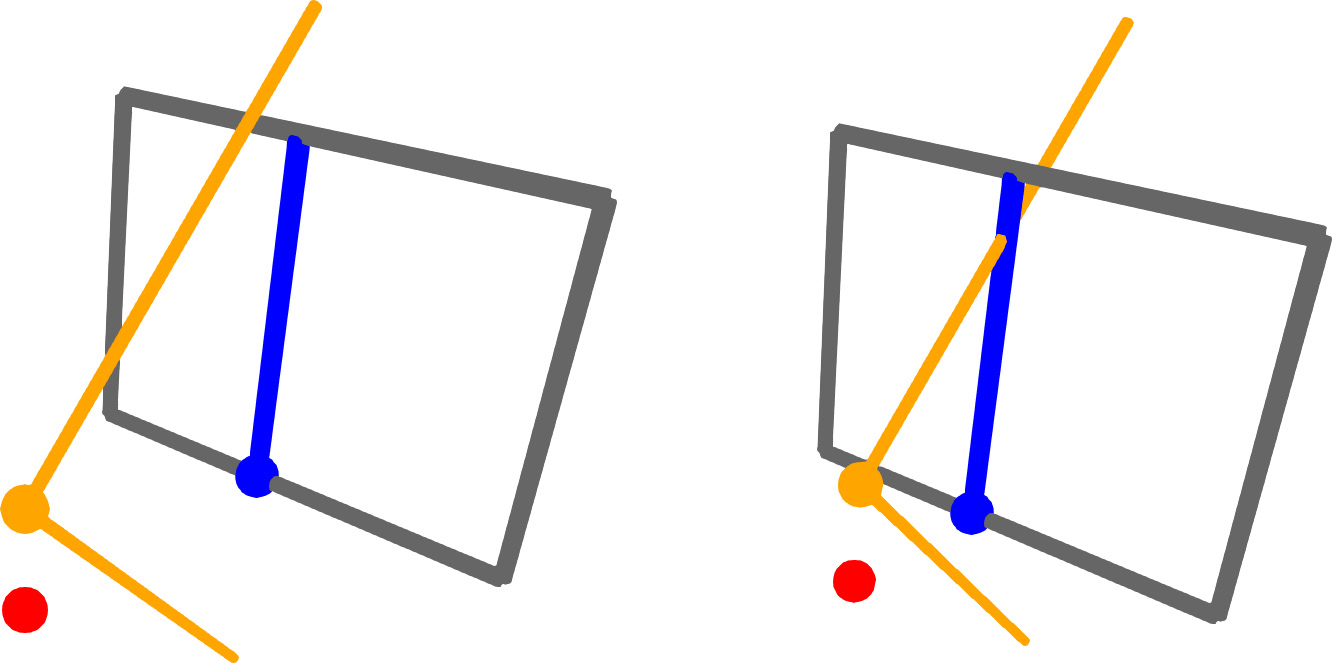}
  \caption{\label{fig:wall} Left: a floor-to-ceiling vertical line (blue) on the wall (black rectangle) is encoded as a 2D point (blue) on the floor plane. The camera in 3D (orange point) is encoded as a 2D point (red) on the floor plane. The encoding is adequate because in this case, the vertical field-of-view angle of the camera, shown by the two orange lines, is large enough to completely cover the vertical line. Right: a case where the camera is too close to the wall such that the vertical line can no longer be completely covered.} 
\end{figure}

To simplify discussions, we define the ROIs to be the walls of the indoor environment only (excluding the floor and the ceiling). Recall that a 360$^{\circ}$ camera's looking directions, in terms of spherical coordinates, span completely horizontally ($0^{\circ}$ to $360^{\circ}$ in azimuth) and nearly completely vertically ($k^{\circ}$ to $(1-k)^{\circ}$ in zenith, $k$ is the "cut-off" vertical angle at the top and bottom nadirs). This means that the visibility of a ceiling-to-floor vertical line (aligned to the z-axis) on a wall w.r.t. a 360$^{\circ}$ camera can be encoded by a point in the x-y plane as long as the 360$^{\circ}$ camera is not too close to the wall - that is, the 360$^{\circ}$ camera either sees the whole vertical line or none of it. Explicitly, we assume the distance of a 360$^{\circ}$ camera to a wall is not smaller than:
\begin{equation}
\label{equ1}
cotangent(FOVy/2) * h, 
\end{equation}
$FOVy$ is the vertical field-of-view angle of the 360$^{\circ}$ camera and $h$ is the bigger value of the camera heights to the floor and to the ceiling. Note that $FOVy$ equals $180^{\circ} - 2k$. See Figure~\ref{fig:wall} for an illustration.

In this way, the visibility problem is simplified as a 2D problem in which the indoor environment is encoded as a 2D piece-wise linear polygon of which each edge corresponds to a wall, and each 360$^{\circ}$ camera is encoded as a 2D point. We denote the former as the "boundary polygon" and the latter as the "camera positions". The visibility problem then requires that all points on the boundary polygon are visible from at least one of the camera positions - i.e., there exist a line in between the boundary point and the camera position that doesn't intersect with any other edges of the boundary polygon.

Same as in previous work~\cite{DG05}, to speed up computations, we uniformly sample the boundary polygon to be a discrete set of "boundary points", $B_i$, $0 \le i < N_b$, $N_b$ is the number of boundary points. We also uniformly sample the interior of the boundary polygon as "interior points", $I_i$, $0 \le i < N_i$, $N_i$ is the number of interior points. {\em We consider these interior points as the candidates of 360$^{\circ}$ camera placements.} We sample the boundary polygon by lengths and sample the interior by grid positions.

To test if a boundary point is visible to an interior point, we use the circular ray sweeping-based algorithm~\cite{asano1985efficient} to tessellate the visible region from an interior point inside the boundary polygon into a set of triangles. We then enumerate boundary points that fall within the triangles of the interior point. A walkthrough of the algorithm is provided in~\cite{Patel20}.

The visibility problem now states that, for every boundary point, it is visible from at least one of the placed 360$^{\circ}$ cameras (which lie on the interior points). In addition to the visibility test mentioned previously, we require that for a placed 360$^{\circ}$ camera to cover a boundary point, two additional tests must be passed: 
\begin{enumerate}
    \item The L2 distance in between should fall within a feasible range. The lower bound of the range can be determined by equation~\ref{equ1} or by a user-specified value (e.g., a 360$^{\circ}$ camera cannot be put too close to a wall due to physical setups). The upper bound (i.e., the maximal visible range of a 360$^{\circ}$ camera) can be determined by referencing the specification of the camera, or, as done in our experiments, empirically determined by physical tests.
    \item The angle between the camera's viewing direction and the boundary point's normal should not be greater than a upper bound. Again, we found the angle upper bound by physical tests.
\end{enumerate}

We now formulate the visibility problem as a linear integer-programming (IP) problem as follows:
\begin{equation}
\begin{aligned}
& \underset{X_i, \ 0 \le i < N_i}{\text{argmin}}
& & \sum{X_i} \\
& \text{subject to}
& & \forall B_j, \ \sum_{k=0}^{n_{B_j}}{X_{j,k}} >=1, 
\end{aligned}
\end{equation}
where $X_i$ denotes a Boolean variable indicating whether a 360$^{\circ}$ camera is placed at $I_i$ (the i-th interior point).  $X_{j,k}$, $0 \le k < n_{B_j}$, enumerate the Boolean variables of potential camera placements at interior points that cover boundary point $B_j$, $n_{B_j}$ is the number of such interior points. In short, the goal is to find a minimal set of 360$^{\circ}$ camera placements such that for every boundary point, at least one of the interior points that cover the boundary point has a camera placed.


\begin{figure}[t]
  \centering
  \includegraphics[width=1\linewidth]{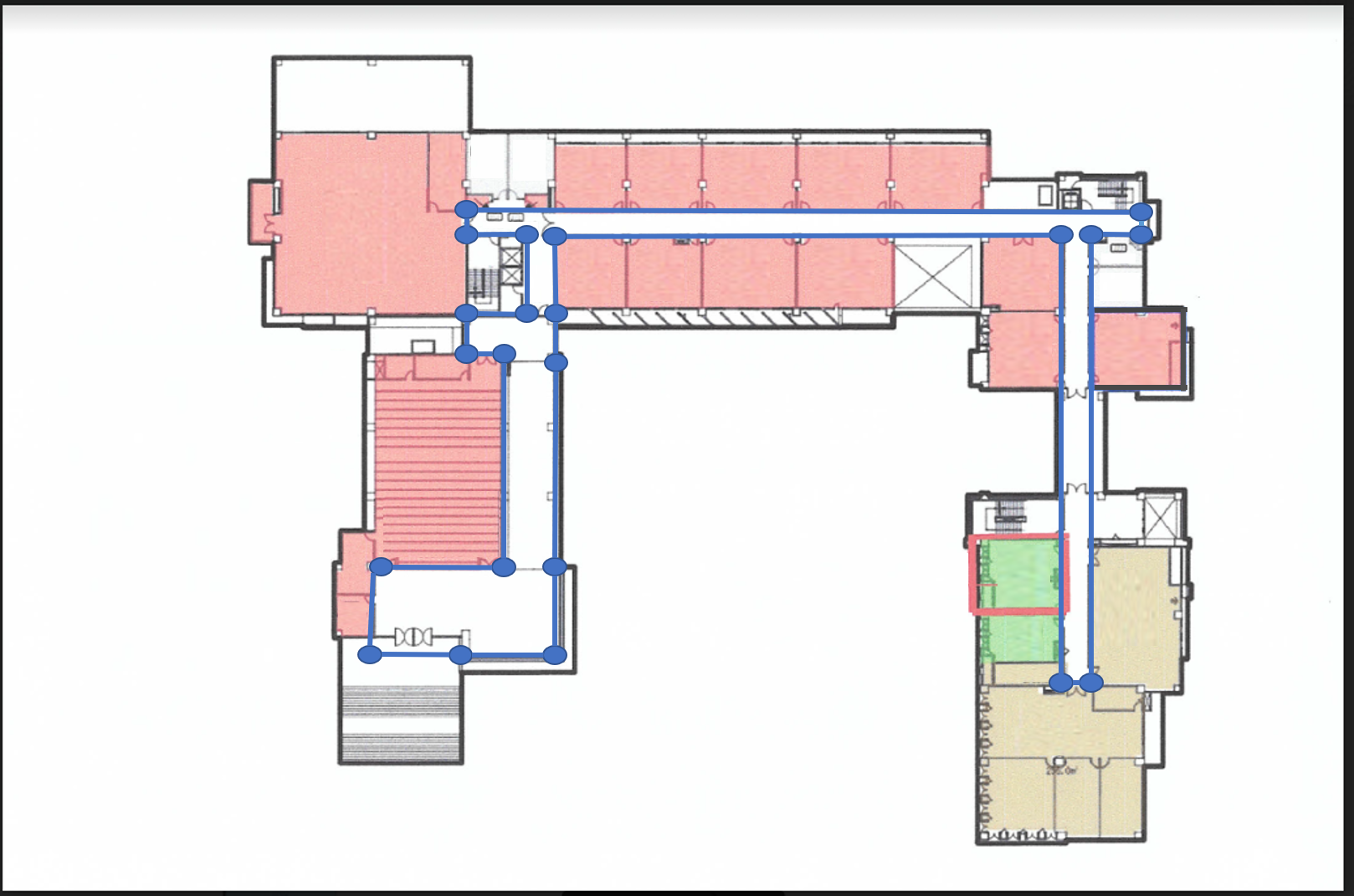}
  \caption{\label{fig:environment} The floorplan of our department that we used to model the problem polygon in our optimization problems.} 
\end{figure}







\section{Results}

 \begin{figure*}[]
  \centering
  \includegraphics[width=1\linewidth]{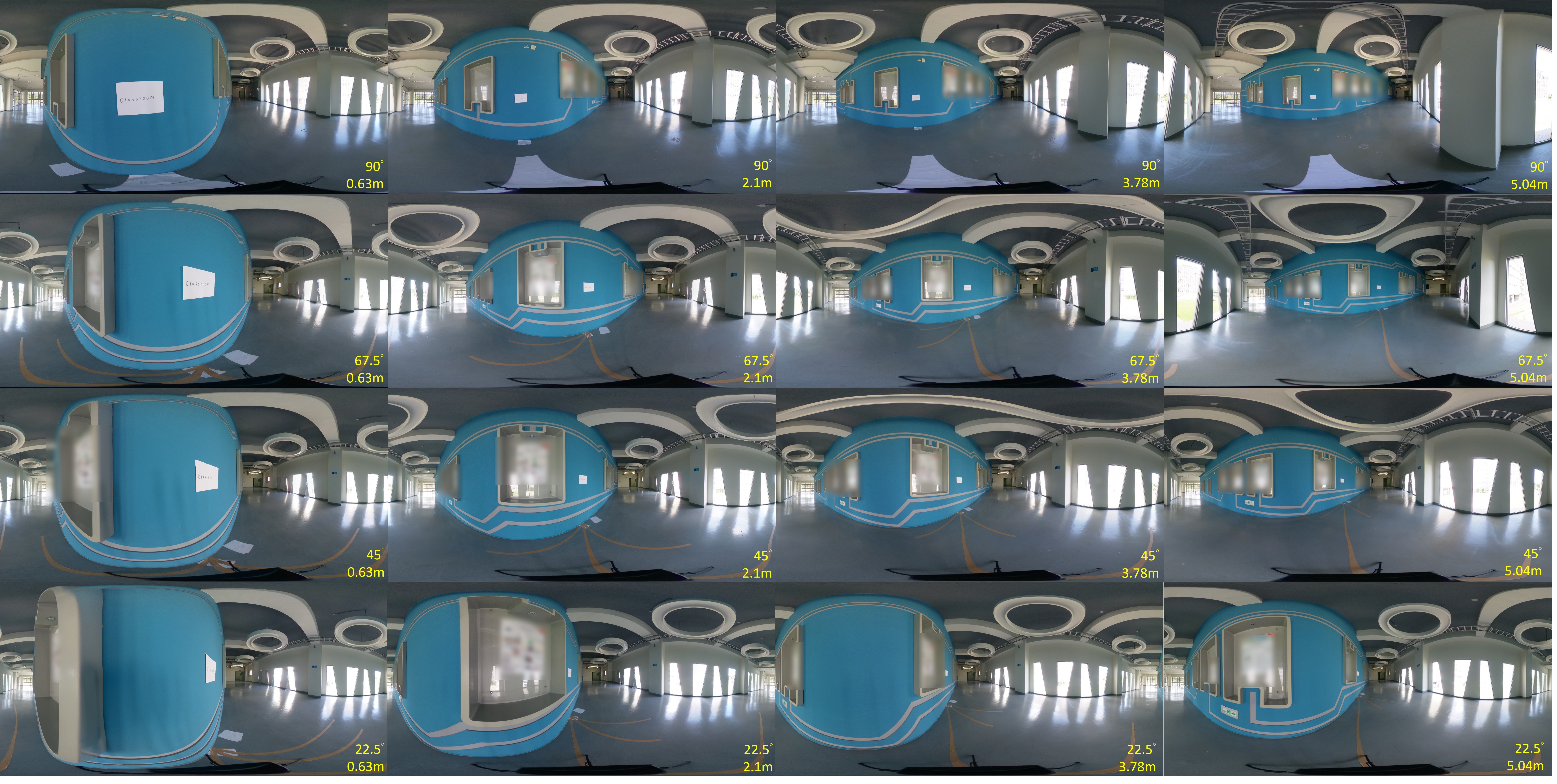}
  \caption{\label{fig:additional} The shoot panoramas which are used to estimate the maximal range and angles of our 360$^{\circ}$ camera (RICOH Theta Z1). In the top-row case, we found that the camera could barely see the "classroom" characters clearly on the A4 paper at about 5.04 meters distance. Therefore, we set 5.04 meters as the camera maximum visibility range in our research. In the second, third, and fourth rows, we show panoramas shoot at different angles to the surface (wall) normal. When camera is at 22.5 $^{\circ}$ to the normal, the characters are too slanted to be seen clearly. Therefore, we choose 45 $^{\circ}$ as the maximum visible angle range.} 
 \end{figure*}

We solved the problem on a desktop computer with Intel i7 6-core 2.60GHZ CPU, NVidia Geforce GTX 1650Ti GPU, and 64GB ram. We used a RICOH Theta Z1 360$^{\circ}$ camera to shoot the panoramas. We model the indoor environment according to the public area (e.g., corridors) of the floor plan of our department (see Figure~\ref{fig:environment}). To empirically find the maximal visible range and the maximal angle (w.r.t. the surface normal) of the camera, we put a paper with printed texts on a wall and shoot numerous panoramas at different distances and angles. The shoot panoramas are shown in figure~\ref{fig:additional}. We found the camera can barely capture characters of font size 118 points printed on a A4 paper at 5.04 meters away and at an angle of 45 degrees. For the minimal range, we find that the camera cannot get closer than 0.61 meters to a wall because the tripod needs space.


\subsection{Optimization results}

\begin{figure*}[t]
  \centering
  \includegraphics[width=1\linewidth]{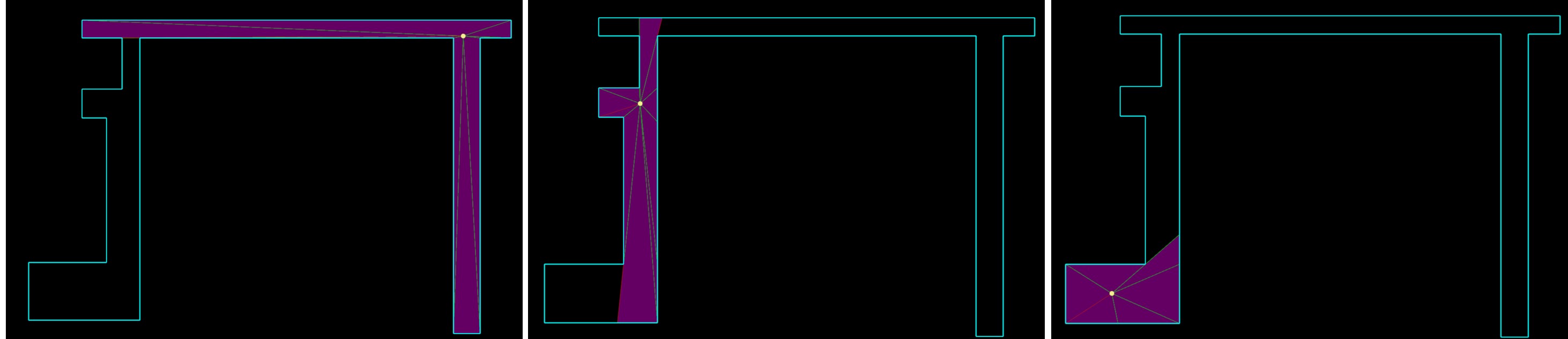}
  \caption{\label{fig:result_nolimit} Coverages of the three 360$^{\circ}$ cameras that together completely cover the whole indoor environment in the non-constrained problem.} 
\end{figure*}

As mentioned previously, we computed three kinds of results: 1) no maximal range nor angle constraints, 2) with maximal range constraint (5.04 meters), and 3) with both maximal range constraint and angle constraint (45 degrees). The results are shown in Figure~\ref{fig:teaser}. As expected, the optimal solutions have more 360$^{\circ}$ camera placements when more constraints are imposed. To show how the cameras together entirely capture the indoor environment, in Figure~\ref{fig:result_nolimit}, we show the coverage areas of the three cameras in the "no-constraint" solution (Figure~\ref{fig:teaser} (a)).

\begin{figure}[t]
  \centering
  \includegraphics[width=1\linewidth]{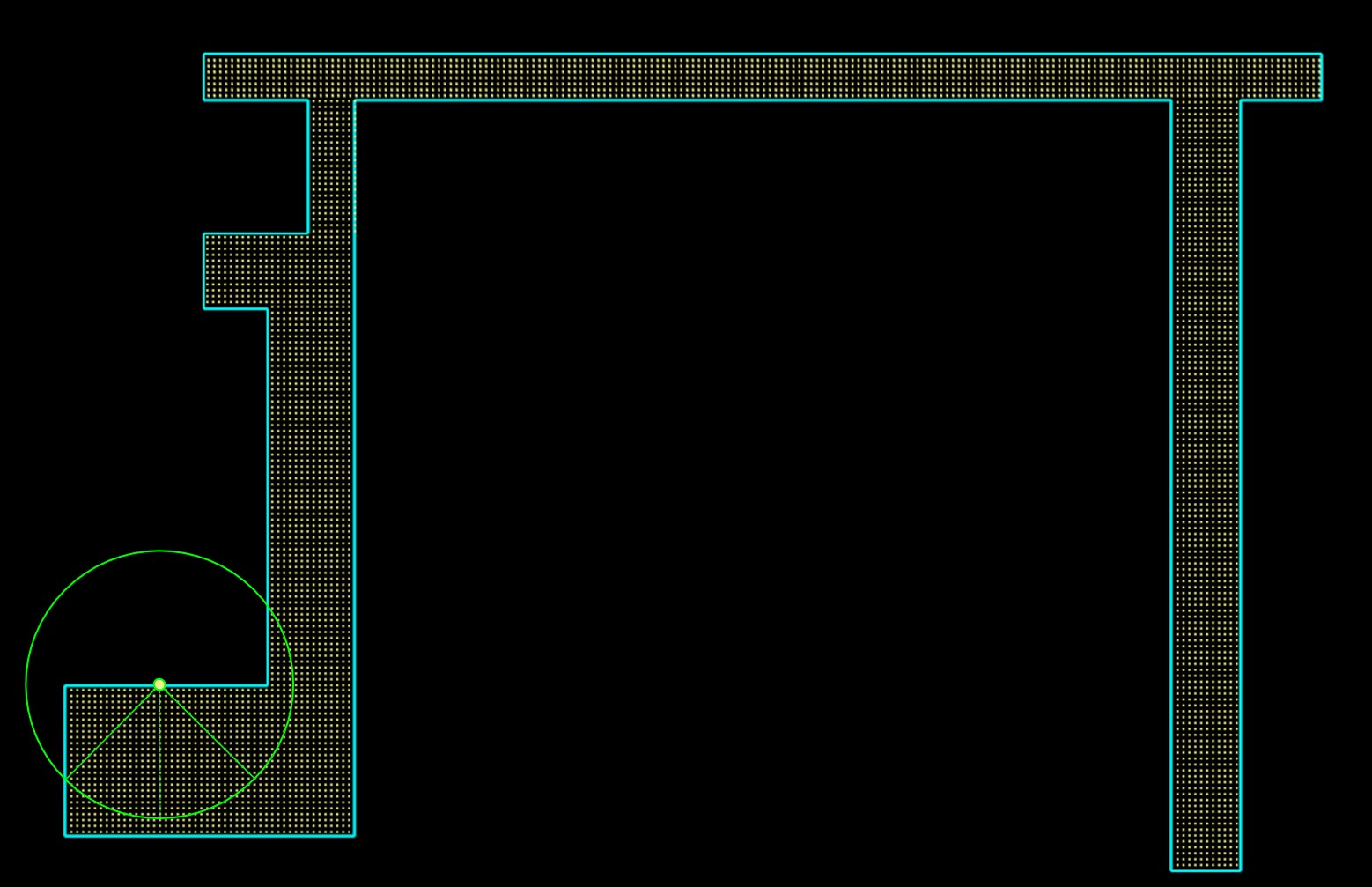}
  \caption{\label{fig:point_range} The sampled boundary and interior points. We also show the set of interior points that cover a boundary point as the ones within the 90-degree quadrant.} 
\end{figure}

We now discuss statistics about the IP problem formulation. For the boundary polygon, we densely sampled 1056 boundary points and 19248 interior points. The sampling is shown in Figure~\ref{fig:point_range}. This means that the visibility matrix (as used in \cite{ghanem2015designing} to describe problem sizes) has about 20.326 million entries, which is bigger than the problems in~\cite{ghanem2015designing}. However, the matrix size alone is not enough to present the sizes of the optimization problems as many pairs of boundary and interior points are mutually invisible to each other. Therefore, we calculate a proxy to the optimization problem sizes as follows. For each boundary point, we count the number of interior points that cover it. We then sum up the counts. The numbers are reported in Table~\ref{tab:table1} first row. Note that the more constraints, the smaller the problem sizes. 

For timing statistics, all the three optimization problems are solved within one second by Gurobi (see Table~\ref{tab:table1} second row). However, we used a Python-based implementation~\cite{Siemering18} of the circular ray sweeping-based algorithm to enumerate visible interior points of a boundary point. For our problem, it took about 26.5 seconds in total to find the boundary points covering every interior points. We expect changing to a C++ implementation with multi-threading should significantly reduce the time.


\begin{table}[]
\centering
\begin{tabular}{|c|c|c|c|}
\hline
\begin{tabular}[c]{@{}c@{}}\end{tabular} & 
\begin{tabular}[c]{@{}c@{}}Fig.~\ref{fig:teaser} (a)\end{tabular} & \begin{tabular}[c]{@{}c@{}}Fig.~\ref{fig:teaser} (b)\end{tabular} & \begin{tabular}[c]{@{}c@{}}Fig.~\ref{fig:teaser} (c)\end{tabular} \\ \hline
Visibility pairs: & 1229768 & 854113 & 295532 \\ \hline
Solutions: & 3 & 18 & 39 \\ \hline
Opt. time: (sec) & 0.6 & 0.33 & 0.076 \\ \hline
\end{tabular}
\caption{Statistics for the three computed solutions in Figure~\ref{fig:teaser}.}
\label{tab:table1}
\end{table}

 \subsection{User studies}
 
\begin{figure*}[t]
 \centering
 \includegraphics[width=1\linewidth]{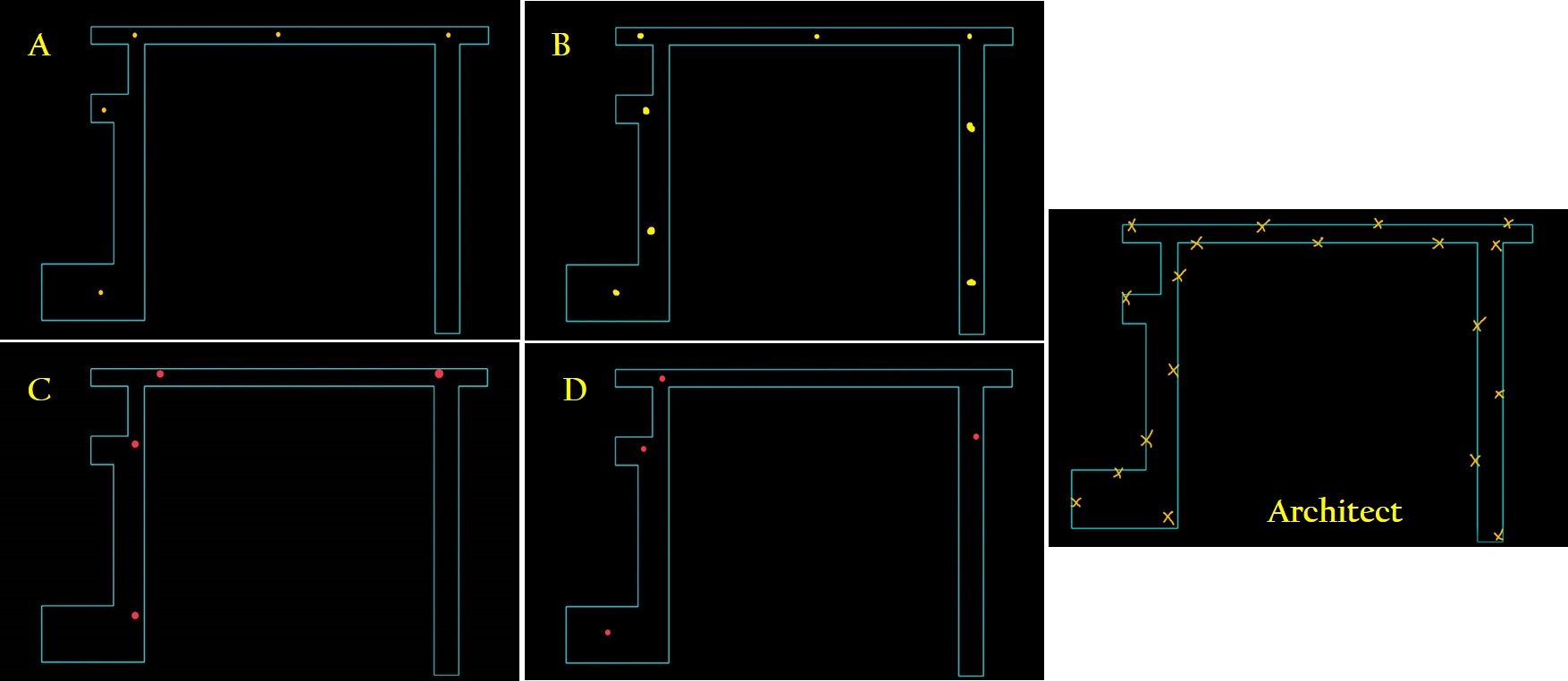}
 \caption{\label{fig:mix_result} Human solutions for the non-constrained problem (Figure~\ref{fig:teaser} (a)).} 
\end{figure*}

\begin{figure*}[t]
 \centering
 \includegraphics[width=1\linewidth]{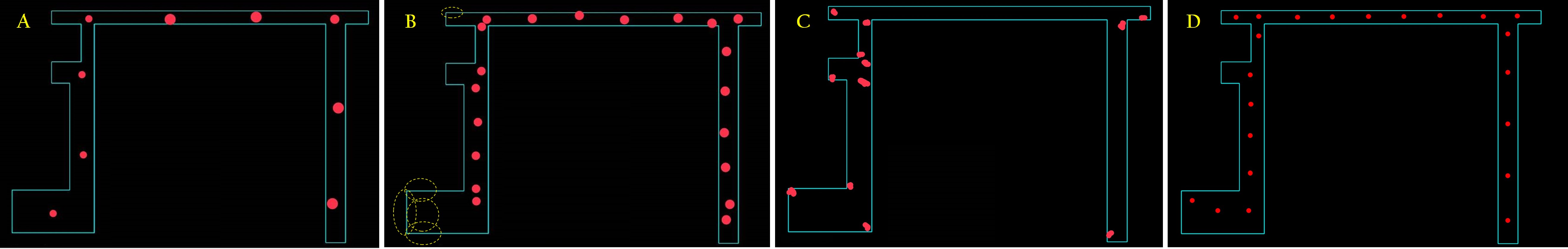}
 \caption{\label{fig:compare_limit} Human solutions for the distance and angle-constrained problem (Figure~\ref{fig:teaser} (c)). In second left, we mark parts of the indoor environment that are missed by the placed cameras.} 
\end{figure*}

We asked several students and one professional architect to solve the same problems without constraints and with maximal range and angle threshold of the cameras. The results are shown in Figure~\ref{fig:mix_result} and Figure~\ref{fig:compare_limit} and reported in Table~\ref{tab:table2}. In short, we find that human users actually have good intuitions on placing 360$^{\circ}$ cameras - when there is no constraints on the ranges nor the angles. However, when range or angle constraints are imposed (Which are necessary in real-world scenarios), humans can no longer give good guesses. The users all took quite some time to decide the solutions. Worse, for the constrained problem, most of the human solutions are actually {\em wrong} as some parts of the indoor environments are not covered by the cameras. The architect gave an interesting (but not optimal) solution for the first problem only.

\begin{table}[]
\centering
\begin{tabular}{|c|c|c|}
\hline
User & \begin{tabular}[c]{@{}c@{}}No constraints\end{tabular} & \begin{tabular}[c]{@{}c@{}}Time (sec)\end{tabular} \\ \hline
A & 5 & 75 \\
B & 8 & 30 \\
C & 4 & 300 \\
D & 4 & 20 \\
Architect & 19 & 60 \\ \hline
& \begin{tabular}[c]{@{}c@{}}W/ dist. and angle constraints\end{tabular} & \begin{tabular}[c]{@{}c@{}}\end{tabular} \\ \hline
A & 9 & 55\\
B & 20 & 79 \\
C & 12 & 242 \\
D & 22 & 286 \\ \hline
\end{tabular}
\caption{The numbers of placed cameras (middle column) and times taken by human users.}
\label{tab:table2}
\end{table}

\section{Conclusion}

In this paper, we leverage modern visibility computation methods in computer graphics, which is critical for the correctness and efficiency of rendering, to solve the problem of placing 360$^{\circ}$ cameras to capture indoor environments in the architecture, engineering, and construction (AEC) industries. We use the circular ray sweeping-based algorithm~\cite{asano1985efficient} to enumerate visible interior points of a boundary point, and formulate the set covering problem as a simple integer-programming (IP) problem that can be readily solved by modern solvers such as Gurobi. With our approaches, globally optimal solutions can be computed very efficiently, making our method suitable for interactive tools.

Our method can only be considered as a first step toward professional solutions for the AEC industries. Therefore, for future work, we aim to expand our methods to accommodate more realistic 3D models of the indoor environments and the 360$^{\circ}$ cameras (for examples, true 3D models instead of approximated 2.5D models) so that cluttered environments can be modeled adequately. We also want to incorporate more real-world issues about 360$^{\circ}$ photography, such as planning the paths of moving the camera/tripod, and deviations of the camera heights, into our computational model.

\section{Acknowledgements}

This work is funded by the Ministry of Science and Technology of Taiwan (110N007  and 111R10286C).


\printbibliography   






\end{document}